# Energy relaxation in edge modes in the quantum Hall effect

Amir Rosenblatt[1], Sofia Konyzheva[1], Fabien Lafont[1], Noam Schiller[1], Jinhong Park[2], Kyrylo Snizhko[1], Moty Heiblum[1], Yuval Oreg[1] and Vladimir Umansky[1]

1. *Braun Center for Submicron Research, Department of Condensed Matter Physics, Weizmann Institute of Science, Rehovot 761001, Israel*
2. *Institute for Theoretical Physics, University of Cologne, Zülpicher Str. 77, 50937 Köln, Germany*

## Abstract

**Studies of energy flow in quantum systems complement the information provided by common conductance measurements. The quantum limit of heat flow in one - dimensional (1D) ballistic modes was predicted, and experimentally demonstrated, to have a universal value for bosons, fermions and fractionally charged anyons. A fraction of this value is expected in non-abelian states; harboring counter-propagating edge modes. In such exotic states, thermal-energy relaxation along the edge is expected, and can shed light on their topological nature. Here, we introduce a novel experimental setup that enables a direct observation of thermal-energy relaxation in chiral 1D edge modes in the quantum Hall effect (QHE). Edge modes, emanating from a heated reservoir, are partitioned by a quantum point contact (QPC) constriction, which is located at some distance along their path. The resulting low frequency noise, measured downstream, allows determination of the 'effective temperature' of the edge mode at the location of the QPC. An expected, prominent energy relaxation was found in hole-conjugate states. However, relaxation was also observed in particle-like states, where heat is expected to be conserved. We developed a model, consisting of distance-dependent energy loss, which agrees with the observations; however, we cannot exclude energy redistribution mechanisms, which are not accompanied with energy loss.**

Pendry [1] was the first to predict a universal upper limit of heat conductance in ballistic one-dimensional modes. This limit holds for any modes of abelian particles, and is independent of particles' exchange statistics [2–4]. Indeed, recent experiments confirmed this upper limit of heat flow with photons [5], phonons [6], electrons [7–11], and fractional charges [10,11]. For modes of non-abelian particles, a fractional upper limit of the thermal conductance is expected, and recently measured in the $\nu = 5/2$ state of the quantum Hall effect (QHE) [12,13]. Moreover, the measured fractional thermal conductance [12] narrowed down the possible topological orders of the non-abelian state [14,15,24,16–23], thus adding information that cannot be obtained by electrical conductance measurements.

A priori, in particle-like QHE states the current and heat flow solely *downstream*, and thus heat is not expected to be lost. However, in electron-hole conjugate states, as well as in non-abelian states, which host counter-propagating modes, heat can be exchanged between the *downstream* and *upstream* modes. This may appear as downstream energy dissipation with a temperature gradients along the edge [3,14,25–28]. Previous studies using quantum dot spectroscopy [29,30] observed energy relaxation at $\nu = 2$; however, as only one of the two co-propagating edge modes was driven far from equilibrium, a strong inter-edge equilibration was suggested as the main relaxation mechanism, while an additional dissipation process remained inconclusive [31–33].

In the following, we chose to coin three useful terms, which will be apparent in the rest of the paper: *relaxation* – *any* redistribution of the energy profile; *equilibration* – a relaxation without energy loss; *dissipation* – relaxation due to energy loss. Due to current conservation, any relaxation process will preserve the low frequency component of the noise ($hf \ll k_B T$, $f$ - frequency, $h$ - Planck constant, $k_B$ - Boltzmann constant and $T$- temperature), along the propagation path.

In our study, we employ a new method to detect energy relaxation in QHE edge modes. A key element, a floating small reservoir (ohmic contact) [9–12,34–36], is heated with its temperature governed by the power dissipated in it. Edge modes, carrying Johnson-Nyquist noise [37,38] with zero net current, are partitioned along the path with a partly pinched QPC, with the resulting noise providing information of the energy relaxation at the location of the QPC.

A dissipation based model, which consists of low-frequency current conservation [39], is found to agree with the data. However, as previous measurements seemed to rule out energy loss in integer modes [9,10,12], energy equilibration without dissipation in our configuration cannot be ruled out.

The experimental setup, shown in Fig. 1a, consists of two regions of high mobility two-dimensional electron gas (light grey) embedded in GaAs-AlGaAs heterostructure, with an electron density $0.9 \times 10^{11}$ cm$^{-2}$ and a mobility of $4.6 \times 10^6$ cm$^2$V$^{-1}$s$^{-1}$. The two regions, separated by a narrow etched trench (dark grey), are bridged by a small, floating, ohmic contact (serving as the heated reservoir, red rectangle). Two current sources, $S_1$ and $S_2$, inject currents (with voltages) $I_1$ ($V_1$) and $I_2$ ($V_2$), which impinge at the floating reservoir. Equilibrating in the reservoir, they raise its temperature from base temperature $T_0$ to $T_m$ [9–12,34–36]. The emerging edge modes, with or without net current, carry current fluctuations, are being partitioned by the QPCs; QPC1 @ 20 µm and QPC2 @ 140 µm. Low frequency voltage fluctuations (~1 MHz) are measured by the amplifier, situated 240 µm away from the heated reservoir.

The reservoir acquires a mean voltage, $V_m = \frac{1}{2}(I_1 + I_2)/G_H$, with $G_H = \nu e^2/h$ is the quantized Hall conductance at filling factor $\nu$. The dissipated power in the reservoir is $\Delta P = (I_1 - I_2)^2/4G_H$. When both currents are equal, $V_m = V_1 = V_2$ and no power is dissipated in the reservoir. Using this structure, $V_m$ and $\Delta P$ can be controlled independently with the two external current sources [35]. A particularly interesting condition, which we exploit in this work is $I_1 = -I_2$, leading to $V_m = 0$, however, with an increased temperature of the reservoir.

The dissipated power in the floating reservoir is evacuated via two means: edge modes and lattice phonons [40–42]. Under steady state conditions the power balance equation gives,

$$\Delta P = \frac{K}{2}(T_m^2 - T_0^2) + \dot{Q}_{\text{el-p}}(T_m, T_0), \quad \#(1)$$

with $KT_m$ is the thermal conductance of edge modes leaving at the reservoir's temperature. The thermal conductance of a single (ballistic) edge mode is a universal number, $\kappa_0 T = \frac{\pi^2 k_B^2}{3h}T$. Note that $\kappa_0$ is independent of the charge and the statistics of the heat carrying quasiparticles [1–4]. The term $\dot{Q}_{\text{el-ph}}(T_m, T_0)$ represents the contribution of the lattice phonons to the dissipated heat, with an increased importance at higher temperature [40–42].

The small heated contact, having capacitance $C$, supports potential fluctuations at frequencies limited by $1/2\pi R_H C$ [36], with $R_H$ the Hall resistance; thus leading to a modified Johnson-Nyquist noise spectrum. In the limit of a small capacitance $\left(2\pi R_H C \ll \frac{h}{k_B T_m}\right)$, the elevated reservoir's temperature is deduced by measuring the low frequency excess voltage noise by the amplifier, $\Delta S_{amp}^V$, being proportional to $\Delta T = T_m - T_0$,

$$\Delta T = \frac{G_H}{k_B} \Delta S_{amp}^V. \quad \#(2)$$

In Fig. 1b we present an example of the measured temperature at $\nu=2$, as function of $I_1$ and $I_2$ (see also Supp. Section I). Two special cases are: (*i*) A diagonal white line, corresponding to $I_1 = I_2$, with $V_m = V_1 = V_2$ and $T_m = T_0$. (*ii*) A diagonal black line, corresponding to $I_1 = -I_2$, with $V_m = 0$, without net downstream current, and $T_m > T_0$. Our main interest here is the latter biasing conditions.

What will the amplifier, located downstream, measure when the current is partitioned previously by a QPC? (*i*) The ubiquitous low frequency shot-noise, at $T=T_0$ [43]. (*ii*) The *partitioned thermal-noise* (emanating from the heated contact at $T_m$, without shot noise). The latter is expected to be sensitive to the energy distribution of the charge modes [44].

To model energy loss in the edge mode, we introduce a fictitious floating contact, placed right before the partitioning QPC (Fig. 1c). The fictitious contact is assumed to be equilibrated with temperature $T_{rel}$. Current conservation along the edge is imposed by demanding that the current entering the floating contact is identical to the current that exits (at low frequencies). In the absence of partitioning, current conservation dictates that the thermal current fluctuations (emanating from the heated floating contact), all propagate towards the amplifier - unaffected by energy loss or redistribution.

Partitioning the edge modes, on the other hand, give access to their relaxation by measuring noise at the downstream amplifier [39,45–47] (see Supp. Section II),

$$\Delta S^V(R) = \frac{k_B}{G_H} \Delta T [R^2 + 2\Theta R(1-R)], \quad \#(3)$$

with $\Delta T = T_\text{m} - T_0$, $R$ the reflection coefficient of the QPC, and $\Theta$ a relaxation parameter, defined as $\Theta = \frac{T_\text{rel} - T_0}{T_\text{m} - T_0}(1+h)$. The parameter $h$ is the "thermal-shot-noise" (recently observed [35,48]), being $h \ll 1$ for small $\Delta T$. The first term in Eq. (3), $\propto R^2$, arises due to the noise generated by heated floating contact and reflected at the QPC. The second term is a result of the temperature difference across the QPC, which generates an additional noise, depending on local temperature of the edge mode and proportional to the shot-noise-like dependence of $R(1-R)$.

The expected dependence of the partitioned noise at $\nu = 2$ on $R$, in the range $0 \leq \Theta \leq 1$, in two extreme cases, is shown in Fig. 1d. The red dots represent the measured noise with QPC1 partitioning (QPC2 fully pinched) - corresponding to unrelaxed edge mode, $\Theta = 1$. The blue dots correspond to noise measured with QPC2 partitioning, but after allowing the edge mode to pass through a massive floating ohmic contact at $T_0$ (QPC1 is fully open and contact grounding removed) - corresponding to a fully relaxed edge mode, $\Theta = 0$ (see Supp. Sections IV and V).

The partitioned thermal noise, by QPC1 and by QPC2, was measured as a function of $\Delta T$, at $T_0 = 15$ mK and $R \approx 0.5$ at a few filling factors (Fig. 2). The full extent of $0 \leq \Theta \leq 1$ is shown in gray. In particle-like states, $\nu = 2$ and $\nu = 1/3$, we find $\Theta \approx 1$ at short propagation distance, with QPC1 partitioning. However, $\Theta \approx 0.5$ was found at the location of QPC2 - suggesting a significant energy relaxation. For the hole-conjugate state, $\nu = 2/3$, which harbors counter-propagating modes, significant relaxation is observed in the nearby location of QPC1. Here, thermal equilibration is expected to take place between the counter-propagating modes [3,25,26,28], bringing energy back to the source, which appears as dissipation. A similar behavior is observed in $\nu=1$, where edge reconstruction is known to take place, giving rise to an underlying $\nu=2/3$ state [49,50]. It is worth noting that the observed linear dependence of the partitioned thermal noise in a wide range of $\Delta T$ (constant relaxation parameter $\Theta$) suggests that the relaxation processes are only weakly dependent on $T_\text{m}$; however, as we show below, it strongly depends on $T_0$ (Fig. 3).

A further confirmation of our theoretical model was tested by measuring the normalized partitioned thermal noise $\Delta S(R)/\Delta S(R=1)$ as a function of $R$ at $T_0 = 15$ mK and $T_0 = 50$ mK (Fig. 3, and Sup. Section VI). The data taken at $T_0 = 15$ mK confirms the one presented in Fig. 3. However, at $T_0 = 50$ mK, severe relaxation is observed already in the location of QPC1 in all the QHE states; being a sign of an additional activated relaxation process. Excitation of phonons, stimulated by their increased occupation at higher temperature, is a possible cause [40–42]. An apparent change in the relaxation rate can be observed near $R \approx 1/2$ in the $\nu = 2/3$ state (and somewhat weaker near $R \approx 2/3$ at $\nu = 1$). In both cases reconstruction with $\nu = 1/3$ edge modes, accompanied by upstream neutral modes, suggests energy exchange between counter propagating modes [51–54].

While the energy dissipative model shown here seems to be consistent with the data, a different model that assumes equilibration without energy loss, was also considered (see Supp. Section III), [36]. However, such equilibration model does not conserve the low frequency noise (without partitioning by a QPC). Alternative models that conserve both total energy and low frequency noise (not developed yet), might also be consistent with the experiment [55].

Here, we studied the evolution of heat propagating in QHE chiral modes. The ballistic nature of modes does not guarantee that their energy distribution remains intact a distance away. We employed a novel technique to study of the evolution of the energy distribution with the propagation length, using a structure that allowed heating a reservoir while maintaining its electro-chemical potential at zero. Two quantum point contacts, placed along the propagation path, were used to partition the heated edge modes, generating partitioned thermal noise. A simple dissipative-based model of the partitioned noise was found to agree with the measured data. Our findings show that energy relaxation takes place even in particle-like states (integer and fractional) as the propagation length exceeds ~100μm. Moreover, the energy relaxation was found to be weekly dependent on the reservoir's temperature, while strongly dependent on the electrons' base temperature - suggesting that stimulated emission of phonons is likely to play a major role in the dissipation process. At the same time a significant relaxation at $\nu = 2/3$ was measured already at low temperatures and at short propagation distance (~20 μm), implying that the energy exchange between the counter-propagating modes is extremely effective.

**Acknowledgements**

We thank Ron Melcer, Yuval Gefen and Yigal Meir for insightful discussions. M.H. acknowledges the partial support of the Israeli Science Foundation (ISF), the Minerva foundation, and the European Research Council under the European Community's Seventh Framework Program (FP7/2007– 2013)/ERC Grant agreement 339070. Y.O. acknowledge the partial support of the ERC under the European Union's Horizon 2020 research and innovation programme (grant agreement LEGOTOP No 788715), the DFG CRC SFB/TRR183, the BSF and NSF (2018643), the ISF (1335/16), and the ISF MAFAT Quantum Science and Technology (2074/19). JP acknowledges funding by the Deutsche Forschungsgemeinschaft (DFG, German Research Foundation) – Projektnummer 277101999 – TRR 183 (project A01). K.S. acknowledges funding by the Deutsche Forschungsgemeinschaft (DFG, German Research Foundation) – Projektnummer 277101999 – TRR 183 (project C01), by the Minerva foundation, and by the German-Israel foundation (GIF). A.R., F.L., S.K. and M.H. designed the experiment. A.R. fabricated the device. A.R. S.K. performed the measurements. A.R. S.K and N.S. did the analysis. A.R., S.K., N.S., J.P., K.S., F.L., Y.O. and M.H. contributed to the theoretical model. V.U. grew the 2DEG heterostructure. All contributed to the write up of the manuscript.

Fig. 1

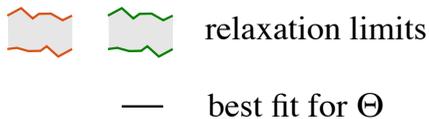 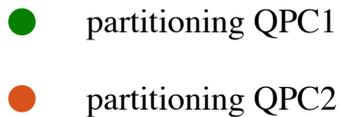
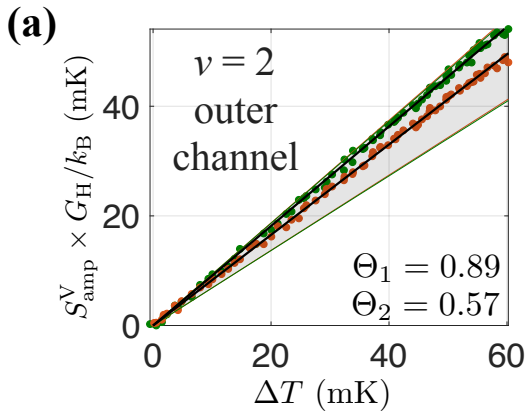 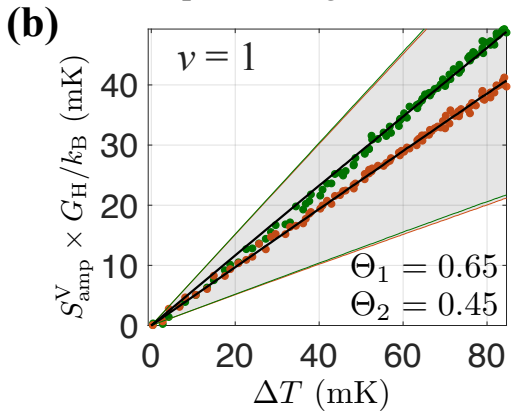
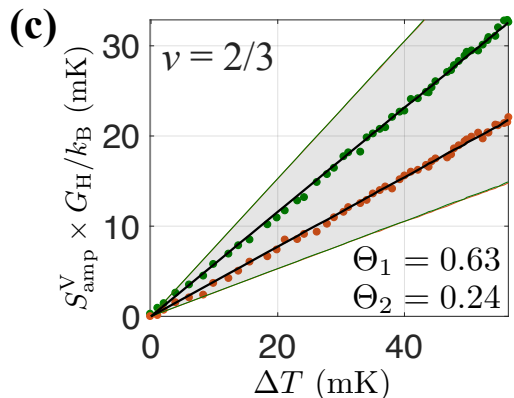 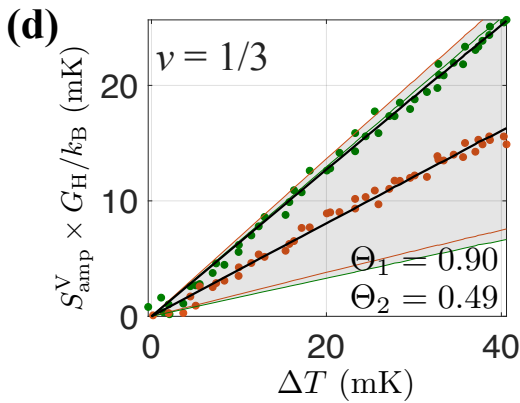

Fig. 2

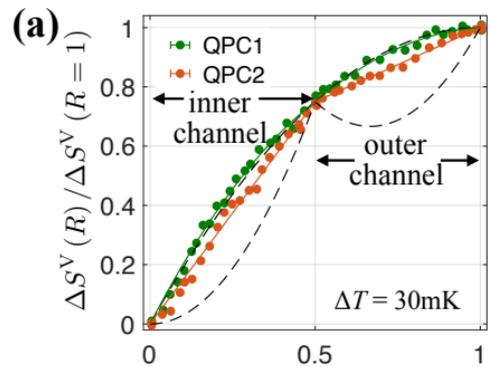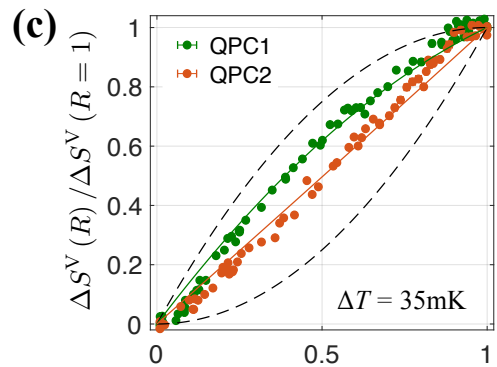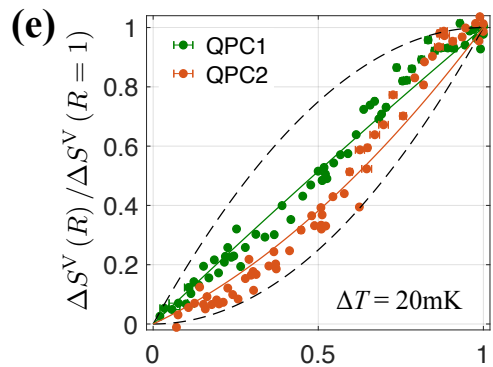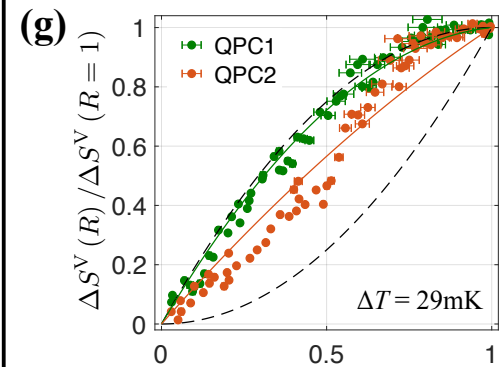
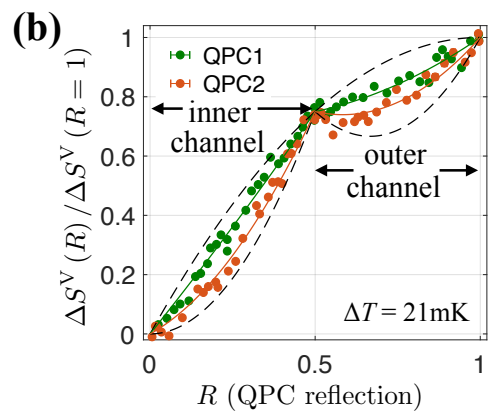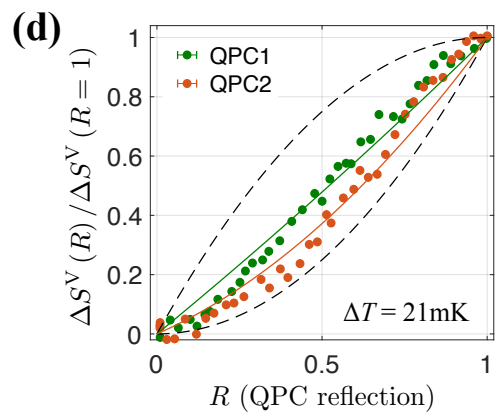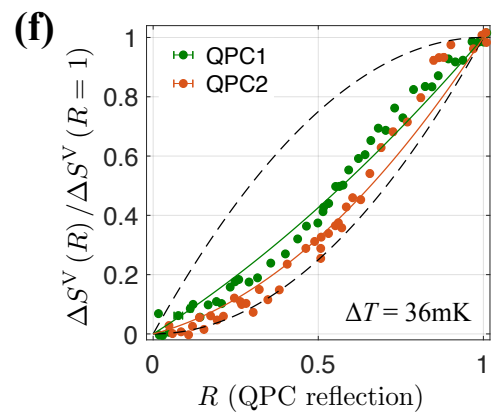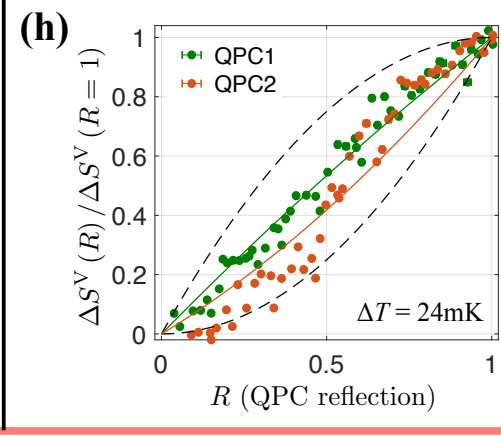

Fig. 3

# Figure captions

**Figure 1. (a)** Scanning electron microscope image of the device: two mesas, separated by an etched groove (dark grey), are connected by a small floating ohmic contact ($10 \times 1.5 \mu m^2$). Two QPCs defined by gates (brown) are placed downstream at 20μm and 140μm away from the floating contact. Two DC sources, $S_1$ and $S_2$, with voltages $V_1$ and $V_2$, and currents $I_1$ and $I_2$ charge the floating ohmic contact, and heat it up. An amplifier, placed 240μm downstream away from the contact, measures the voltage noise. **(b)** Measured temperature of the floating contact, $T_m$, as function of $I_1$ and $I_2$, at $v = 2$ and $T_0 = 15$mK. The temperature is extracted from the measured Johnson-Niquist noise, when both QPCs are fully pinched. **(c)** Schematic description of the circuit, which contains a fictitious floating contact inducing dissipation. Current flows from two sources with voltages $V_1, V_2$, and temperature $T_0$ to the small floating contact. The floating contact is at voltage $V_m$ and temperature $T_m$. Current leaves the floating contact via two arms - one is partitioned by a QPC en-route to the Amplifier contact. A fictitious floating contact, with temperature $T_{rel}$, representing the effective local temperature, is placed right before the QPC. The partitioned current is reflected toward the amplifier. **(d)** Normalized partitioned thermal noise as function of the reflection coefficient of the QPC, $R$, for different values of the relaxation parameter $\Theta$ at a constant $\Delta T$. Red dots: Data measured with partitioning by QPC1 at $v = 2$. Blue dots: Data measured with partitioning by QPC2 at $v = 2$ after cooling down the edge modes.

**Figure 2.** Partitioned thermal noise as function of temperature difference $\Delta T$ at a constant reflection coefficient $R \approx 1/2$, at $T_0 = 15$mK. Green circles – partitioning by QPC1 while keeping QPC2 fully pinched, Orange circles- partitioning by QPC2 while keeping QPC1 fully pinched. Grey region – limits of the theoretical model. Lower limit corresponds to full relaxation, $\Theta = 0$ and upper limit corrsponds to $\Theta = 1$. Black lines – linear fit for the relaxation parameter, $\Theta$, using Eq. 3. **(a)** $v = 2$ outer channel, **(b)** $= 1$, **(c)** $v = 2/3$, **(d)** $v = 1/3$.

**Figure 3.** Partitioned thermal noise as a function of the reflection coeficient at a constant temperature difference. Green circles – partitioning QPC1 while keeping QPC2 fully pinched; Orange circles - partitioning QPC2 while keeping QPC1 fully pinched; Dashed black lines – limits of the theoretical model, lower limit corresponds to full relaxation $\Theta = 0$ and upper limit corrsponds to $\Theta = 1$. Orange/green solid line – best fit of the relaxation parameter, $\Theta$, using Eq. 3. A sketch of the edge channel is plotted for each filling factor, with solid arrow indicating a downstream charge mode and dashed arrow indicating an upstream neutral mode. **(a,c,e,g)** base temperature $T_0 = 15$ mK. **(b,d,f,h)** base temperature $T_0 = 50$ mK. **(a,b)** $v = 2$. **(c,d)** $v = 1$. **(e,f)** $v = 2/3$. **(g,h)** $v = 1/3$.

# Supplementary Materials

Amir Rosenblatt, Sofia Konyzheva, Fabien Lafont, Noam Schiller, Jinhong Park, Kyrylo Snizhko, Moty Heiblum, Yuval Oreg and Vladimir Umansky

*Department of Condensed Matter Physics,*

*Weizmann Institute of Science, Rehovot 761001, Israel*

## I. Measuring $T_m$ and the quantum limit of heat flow

With the two QPCs are fully pinched, the voltage noise was measured at the amplifier's contact as function of the injected currents, $I_1 = -I_2$, keeping $V_m = 0$. In Fig. S1a, we plot the temperature $T_m$, rendered from the measured noise, exploiting Eq. 2. In Fig. S1b we plot $T_m$ as a function of the dissipated power $\Delta P$ in all the filling factors. The quantum limit of heat is determined from the expression of the power balance:

$$\Delta P = \frac{K}{2}(T_m^2 - T_0^2) + \beta_{\text{el-ph}}(T_m^5 - T_0^5), \qquad (S1)$$

with $KT_m$ ($KT_0$) the thermal conductance of edge modes leaving the reservoir (leaving grounded contacts). The thermal conductance of a single (ballistic) edge mode is a universal number, $KT = \kappa_0 T = \frac{\pi^2 k_B^2}{3h}T$. The heat carried by the phonons is expected to have a $T^5$ dependence, and $\beta_{\text{el-ph}}$ is the electron-phonon coupling constant, which depends on the volume of the heated reservoir [1–3]. The heat power measurements for $\nu = 3$, 2 and 1 (Fig. S1b, circles) agrees with the quantum heat power with $6\kappa_0, 4\kappa_0$ and $2\kappa_0$, for the two-arm device, respectively (Fig. S1b, black curves) with $\beta_{\text{el-ph}} = 7$ nW/T$^5$. Heat power measurements for the fractional $\nu = 1/3$, agrees with $2\kappa_0$ as expected. At the hole-conjugate state $\nu = 2/3$, one should expect zero heat conductance at long distance, due to energy transfer between counter-propagating edge modes. These results, reproduce the quantum limited heat flow measured before [4–6].

## II. Theoretical model: dissipation

Our model attempts to describe the experimental results for integer quantum Hall states with minimal physical assumptions. Current continuity at low frequencies is assumed; namely, no charge accumulates anywhere along the edge [7]. A fictitious floating contact, placed right before the QPC models the dissipation (Fig S2). From here, we follow standard methods of calculating zero-frequency fluctuations in mesoscopic circuits [8–10].

The fictitious floating contact is assumed to have temperature $T_{\text{rel}}$, corresponding to the effective (cooled) temperature of the mode at the point of partitioning. Hence, this temperature spans the range $T_0 \leq T_{\text{rel}} \leq T_m$, with the lower and upper bounds representing full and no relaxation, respectively. No explicit assumptions are made as to the mechanism leading to relaxation, but as noted in the main text, the prominence of relaxation at higher base temperatures points at electron-phonon interactions as a reasonable culprit.

We begin with the integer state $\nu = 1$; generalization to higher filling factors is straightforward. The currents at our real, small floating contact, $I_m$; at the fictitious floating contact, $I_C$; and at the amplifier, $I_{\text{Amp}}$, are given by the equations

$$I_m = G_1(V_m - V_1) + G_2(V_m - V_2) + \delta I_m, \tag{S2a}$$

$$I_C = G_1(V_C - V_m) + \delta I_C, \tag{S2b}$$

$$I_{\text{Amp}} = G_1(V_{\text{Amp}} - RV_C) + \delta I_{\text{Amp}}. \tag{S2c}$$

Here, $G_i$ is the conductance in the $i^{\text{th}}$ side of the mesa (we will later take $G_1 = G_2 = G_H$), $R$ is the QPC's reflection coefficient, and $\delta I_j$ describes intrinsic current fluctuations at contact $j$ in the absence of voltage fluctuations [7]. Subtracting the average value of these three currents, we find,

$$\Delta I_m = (G_1 + G_2)\delta V_m + \delta I_m, \tag{S3a}$$

$$\Delta I_C = G_1(\delta V_C - \delta V_m) + \delta I_C, \tag{S3b}$$

$$\Delta I_{\text{Amp}} = -G_1 R \delta V_C + \delta I_{\text{Amp}}, \tag{S3c}$$

where $\Delta I_j \equiv I_j - \langle I_j \rangle$ and $\delta V_j \equiv V_j - \langle V_j \rangle$. We emphasize that only the floating contacts have non-zero voltage fluctuations, since all other reservoirs are kept at constant chemical potential. Hence, only $\delta V_m$, $\delta V_C \neq 0$. The value of the voltage fluctuations is determined by our assumption that no charge accumulates at the floating contacts, i.e. $\Delta I_m = \Delta I_C = 0$ [7]. Thus, Eq. S3 directly gives,

$$\Delta I_{\text{Amp}} = R \frac{G_1}{G_1 + G_2} \delta I_m + R \delta I_C + \delta I_{\text{Amp}} . \tag{S4}$$

The voltage noise fluctuations measured at the amplifier are defined via the correlation function, $S^V \equiv \frac{1}{G_1^2} \langle \Delta I_{\text{Amp}} \Delta I_{\text{Amp}} \rangle$,

$$G_1^2 S^V = R^2 \left[ \left( \frac{G_1}{G_1 + G_2} \right)^2 S^{\text{eq}}_{\text{m,m}} + 2 \frac{G_1}{G_1 + G_2} S^{\text{eq}}_{\text{m,C}} + S^{\text{eq}}_{\text{C,C}} \right]$$
$$+ 2R \left[ \frac{G_1}{G_1 + G_2} S^{\text{eq}}_{\text{m,Amp}} + S^{\text{eq}}_{\text{C,Amp}} \right] + S^{\text{eq}}_{\text{Amp,Amp}} , \tag{S5}$$

where we define $S^{\text{eq}}_{i,j} \equiv \langle \delta I_i \delta I_j \rangle$, using the zero-frequency limit of the canonical formula for noise correlations [10,11],

$$S^{\text{eq}}_{ij}(\omega) = \frac{e^2}{h} \sum_{k,l} \sum_{m,n} \int dE \, A^{mn}_{kl}(i; E, E + \hbar\omega) A^{nm}_{lk}(j; E + \hbar\omega, E)$$
$$\times \{ f_k(E)[1 - f_l(E + \hbar\omega)] + f_l(E + \hbar\omega)[1 - f_k(E)] \} . \tag{S6}$$

Here, $k, l$ are summed over terminals, $m, n$ are summed over channel numbers with values between 1 and the filling factor (and are hence trivial for filling factor 1). $f_k(E) = \frac{1}{e^{\frac{E - \mu_k}{k_B T_k}} + 1}$ is the Fermi-Dirac distribution of the electrons that exit contact $k$. $A$ is determined by the scattering matrix of the problem via,

$$A^{mn}_{kl}(i; E, E') = \delta_{ik} \delta_{il} \delta_{mn} - \sum_p s^\dagger_{ik;mp}(E) s_{il;pn}(E') , \tag{S7}$$

where $s_{ik;mn}(E)$ denotes the scattering matrix element at energy $E$, relating channel $m$ in contact $i$, to channel $n$ in contact $k$. Performing explicit calculations with all terms of this form in Eq. S5, we obtain,

$$S^V = 2 \frac{k_B}{G_1} \left[ \frac{G_2}{G_1 + G_2} R^2 (T_m - T_0) + R(1 - R)(T_{\text{rel}} - T_0) \left( 1 + h\left(\frac{T_{\text{rel}}}{T_0}\right) \right) + 2 T_0 \right] , \tag{S8}$$

where $h\left(\frac{T_{\text{rel}}}{T_0}\right)$ is the thermal-shot-noise obtained from partitioning channels with different temperatures [12,13], given by the expression

$$h\left(\frac{T_{\text{rel}}}{T_0}\right) = \frac{1}{k_B(T_{\text{rel}} - T_0)} \int dE \left(\frac{1}{e^{\frac{E}{T_{\text{rel}}}} + 1} - \frac{1}{e^{\frac{E}{T_0}} + 1}\right)^2. \tag{S9}$$

It is worth noting at this stage that for $R = 0, 1$, the effective local temperature $T_{\text{rel}}$ drops entirely out of the expression, consistent with the expectation that temperature changes along the edge do not affect low-frequency noise in the absence of partitioning.

Focusing exclusively on the excess noise, $\Delta S^V = S^V - S^V(T_m = T_0)$, with $G_1 = G_2 = G_H$, and $\Theta \equiv \frac{T_{\text{rel}} - T_0}{T_m - T_0}\left(1 + h\left(\frac{T_{\text{rel}}}{T_0}\right)\right)$, the excess noise reduces to the rather compact expression used in Eq. 3,

$$\Delta S^V = \frac{k_B}{G_H}[R^2 + 2\Theta R(1 - R)](T_m - T_0). \tag{S10}$$

It is worth dwelling on the nature of this thermal shot noise. The asymptotic values of $h\left(\frac{T_{\text{rel}}}{T_0}\right)$, as described in Eq. S9, are: $\lim_{T_{\text{rel}} \to T_0} h\left(\frac{T_{\text{rel}}}{T_0}\right) \propto (T_{\text{rel}} - T_0)$, and $\lim_{\frac{T_{\text{rel}}}{T_0} \to \infty} h\left(\frac{T_{\text{rel}}}{T_0}\right) \approx 0.38$. This means that for a small $\Delta P$ this $h$ term can entirely be neglected; yet, even at high $T_m$, its contribution does not dominate the measured noise.

The constraint $T_{\text{rel}} \leq T_m$ dictates $\Theta \leq 1 + h\left(\frac{T_m}{T_0}\right)$, with a maximum of $\Theta \approx 1.38$. In particular, the thermal shot noise contribution explicitly leads to $\Theta > 1$ for no relaxation. We did not find values of $\Theta$ that are larger than unity, hinting that all measured configurations exhibit some relaxation.

Generalization of the result in Eq. S10 to integer filling factors other than 1 is rather straightforward. For an integer filling factor $\nu$, we have in total $\nu$ edge channels, and we assume that each channel $1 \leq n \leq \nu$ is described by its own effective temperature $T_{\text{rel},n}$ and its own reflection coefficient $0 < R_n < 1$, with the total reflection coefficient of the QPC determined by $R = \frac{1}{\nu}\sum_{n=1}^{\nu} R_n$. Carefully repeating the above steps, we arrive at the general formula,

$$\Delta S_\nu^V = 2\frac{k_B \nu}{G_H}\left[\sum_{n=1}^{\nu}\left(1 - \frac{1}{2\nu}\right)R_n^2 - \frac{1}{\nu}\sum_{n'>n=1}^{\nu} R_n R_{n'} + \sum_{n=1}^{\nu} \Theta_n R_n(1 - R_n)\right] \times (T_m - T_0), \tag{S11}$$

where once again, $\Theta_n \equiv \frac{T_{\text{rel},n}-T_0}{T_m-T_0}\left(1 + h\left(\frac{T_{\text{rel},n}}{T_0}\right)\right)$.

It is important to note that mode-dependent quantities are not mutually independent. Since the QPC partitions channels from the inside to outside, at any given point there is only one channel that is not fully reflected or fully transmitted. Thus, for any given value of $R = \frac{1}{\nu}\sum_{n=1}^{\nu} R_n$, all values of $R_n$ are well defined. Additionally, as argued above and readily apparent from Eq. S11, the measured noise is independent of all effective temperatures $T_{\text{rel},i}$ of the modes that are fully reflected or transmitted. Thus, the only free parameter in Eq. S11 is $T_{\text{rel},j}$ for the single partitioned channel $j$. In particular, at $R = \frac{q}{\nu}$ at filling factor $\nu$, which corresponds to $q$ fully transmitted channels and $\nu-q$ fully reflected channels, we obtain,

$$\frac{\Delta S^V\left(R=\frac{q}{\nu}\right)}{\Delta S^V(R=1)} = \left(\frac{q}{\nu}\right)^2 + 2\left(\frac{q}{\nu}\right)\frac{\nu-q}{\nu}. \tag{S12}$$

These specific values fit our data very well for $\nu = 2, 3$ (see Fig. 3 and Fig. S5), thus providing strong support for our model.

### III. Theoretical model – equilibration

In parallel to the model described in the previous section, further efforts were made to model energy relaxation through redistribution of energies without loss. Our assumption was that this redistribution eventually arrives at the thermodynamic limit of an equilibrium Fermi-Dirac distribution. Such a distribution at temperature $T$ and no bias voltage corresponds to a frequency-dependent noise correlation function of

$$S_T(\omega) = G_H \hbar \omega \coth\left(\frac{\hbar\omega}{2k_B T}\right), \tag{S13}$$

giving zero-frequency noise of $S_T(\omega = 0) = 2k_B G_H T$, and a total energy of $\int d\omega S_T \, \omega \propto (k_B T)^2$. As such, both the zero-frequency noise and total energy are mutually determined by a single parameter, and cannot be separately tuned to any pair of desired values. Consequently, by assuming an energy conserving Fermi-Dirac equilibrium, we explicitly broke zero-frequency noise conservation.

This assumption was ultimately not borne out in the data. As such, a model assuming equilibration along the edge towards a Fermi-Dirac distribution, along the lines of Ref. [14], was insufficient in explaining our measurements. In particular, such a model fails to recreate the results of Eq. S12, which serve as a stark feature of the measurement and are obtained through the zero-frequency conservation assumed in Supp. Section II.

### IV. Measurement methodology of partitioned thermal noise

This section demonstrates the methodology of how to determine the thermal relaxation. Using the device presented in Sup. figure 3a, DC power $\Delta P$ is applied while $V_m = 0$, in order to heat the small floating contact but keep its potential at zero. The hot edge modes leaving the small floating contact, encounter two QPCs: QPC1 at distance 20μm away from the small floating contact and QPC2 at 140μm away. On each QPC we can partition the hot edge - flowing on one side, with the grounded edge at $T_0$ - flowing on the other side, resulting partitioned thermal noise, free of any shot-noise.

First, the reflected differential conductance of each QPC is measured with small AC voltage $\sim 0.5\mu V_{RMS}$ such that it has negligible heating. The measurement was done both with and without DC heating and will denote it as an error in the coming plots. We note that the reflected differential conductance of either QPCs has almost no dependence on the heating. The reflected differential conductance of each QPC is measured while scanning the QPC gate voltage, as the other QPC was fully pinched (Sup. figure 3b). Next, we applied constant DC heating power $\Delta P$ raising its temperature to some $T_m$, while keeping $V_m = 0$. The partitioned thermal noise was measured at the amplifier as function of the split-gate voltage $V_{QPC}$ of each QPC (Sup. figure 3c).

In order to determine the thermal relaxation, we plot the normalized partitioned thermal noise as function of the reflection coefficient $R$, in Sup. figure 3d. The upper black curve indicates the non-relaxed limit ($\Theta = 1$), the lower black curve indicates the fully relaxed limit ($\Theta = 0$, for a single edge mode) and the dashed black curves indicate the fully relaxed limit for two edge modes ($\Theta = 0$, for each edge mode).

### V. Testing the full relaxation limit

In this section we probe the partitioned thermal noise of a fully relaxed edge mode. The edge mode is cooled down by entering into a massive floating contact, strongly coupled to the lattice temperature $T_0$ with electron-phonon interactions. The electron-phonon interaction is proportional to the volume of the contact. The massive ohmic contact volume is $10 \times 60 \times 0.2 \mu m^3$, which is ~40 times bigger than the small floating contact. In addition, it is connected to a large metallic pad ($80 \times 80 \mu m^2$), also contributing to thermalization. The RC cutoff frequency of such thermalizing contact is estimated to be in the ~ 100MHz range, which is much smaller than the temperature cutoff (~ GHz range) and much larger than the measuring frequency ~ 1MHz.

The partitioned thermal noise was measured at $\nu = 2$ in two cases: (1) partitioning at QPC1, keeping QPC2 fully pinched (Fig. S4a); (2) keeping QPC1 open, allowing the channel to pass through the massive floating contact and partitioning at QPC2 (Fig. S4c). In case (1), the partitioned thermal noise (Sup. figure 4b), corresponds to non-relaxed edge modes. In case (2), the partitioned thermal noise corresponds to a fully relaxed single edge mode (Sup. Fig. 4d), as expected by the equilibration of the two edge modes inside the massive floating contact.

Note that the same amount of low frequency noise was measured when QPC2 is fully pinched, without any difference whether QPC1 is open or close, confirming conservation of low frequency current on the massive floating contact.

## VI. Thermal relaxation of edge modes vs. reflection coefficient

Following the measurement procedure presented in Sup. section IV, a constant DC heating $\Delta P$ applied while keeping $V_m = 0$, raising the small floating contact's temperature to some $T_m$. The partitioned thermal noise is measured as function of the reflection coefficient of the QPC. The normalized noise $\Delta S(R)/\Delta S(R = 1)$ is presented in Fig. S5 for particle-like states ($\nu = 3, 2, 1/3$) and in Fig. S6 for hole-conjugate states ($\nu = 1+$ at 3.2T, the low field side of the $\nu = 1$ plateau; $\nu = 1-$ at 3.75T, the high field side of the $\nu = 1$ plateau; and $\nu = 2/3$) together with a fit of the dimensionless relaxation parameter $\Theta$ for each edge mode, such that:

$$\frac{\Delta S^V(R)}{\Delta S^V(R = 1)} = R^2 + 2\Theta R(1 - R), \qquad (S14)$$

where $0 < \Theta < 1 + h\left(\frac{T_m}{T_0}\right)$. The limit $\Theta = 0$ corresponds to full relaxation (lower black dashed curve); the limit $\Theta = 1 + h\left(\frac{T_m}{T_0}\right)$ corresponds to no relaxation (upper black dashed curve); and the limit $\Theta = 1$ corresponds to no relaxation with zero thermal-shot-noise contribution $h = 0$ (upper black solid curve). A model for multichannel case (such as in $\nu = 2$ and 3) is found in Sup. section II. Noise measured with partitioning QPC1, placed 20μm away from the small floating contact, plotted in green circles. Noise measured with partitioning QPC2, placed 140μm away from the small floating contact, plotted in orange circles.

Filling factor $\nu = 1$ is known to experience edge reconstruction, giving rise to an underlying $\nu = 2/3$ state [15,16]. This is also suggested from our measurements, having pronounce relaxation at $\nu = 1$ (Fig. S6), growing stronger at the high field side of the $\nu = 1$ plateau, where the reconstruction is expected to appear [15].

### VII. Device fabrication

The GaAs/AlGaAs heterostructure used in the report (#C2-287), has electron density $n = 0.90 \times 10^{11} \text{cm}^{-2}$ and dark mobility $\mu = 4.6 \times 10^6 \text{cm}^2\text{V}^{-1}\text{s}^{-1}$, with the 2DEG buried 1280Å below the surface. For the ohmic contacts, evaporation of the sequence: Ni 50Å, Au 1900Å, Ge 1000Å, Ni 710Å, Au 150Å, followed by annealing at 120°C for 120 sec, followed by 360°C for 30 sec, and 440°C for 50 sec. For all the metallic gates on the device, such as QPCs, we evaporated of Ti 50Å followed by Au 150Å, employing lift-off techniques.

### VIII. Calibration of the amplifier

We calibrated the amplifier using two methods: the first using shot-noise at $\nu = 2$ (Fig. S7a). The outer channel was partitioned and the noise was measured as function of the current (injecting two currents with the same polarity, $I_1 = I_2$), leaving the small floating ohmic contact at a finite voltage and zero power dissipation. We used the ubiquitous shot-noise equation [11]:

$$\Delta S^{\mathrm{I}} = 2eIR(1-R)\left[\coth(x) - \frac{1}{x}\right], \#(S15)$$

where $x = \frac{eI}{2k_{\mathrm{B}}TG_{\mathrm{H}}}$. We fitted both gain and electron temperature, gain=8.8 and $T = 15$mK. Another calibration method used the Johnson-Nyquist noise [17,18], $S^{\mathrm{I}} =$ Gain x $4k_{\mathrm{B}}TG_{\mathrm{H}}$, at a few fridge base temperature, ranging from $T = 15$mK to $T = 300$mK, where $G_H = 2e^2/h$ at $\upsilon = 2$ (Fig. S7b). The fitted Gain=$8.9 \pm 0.2$ - in agreement with the previous shot-noise fit.

# Supplementary Figures

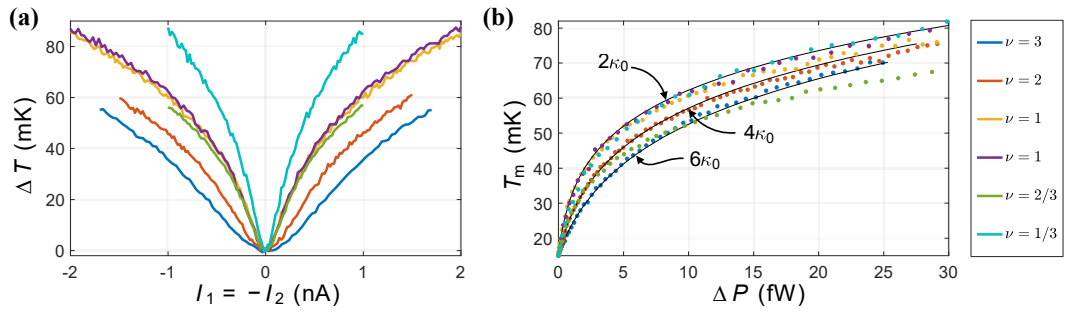

*Figure S1*

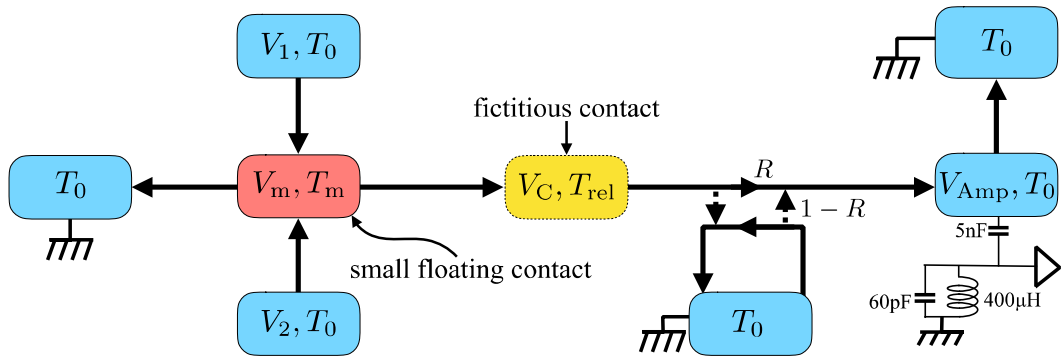

*Figure S2*

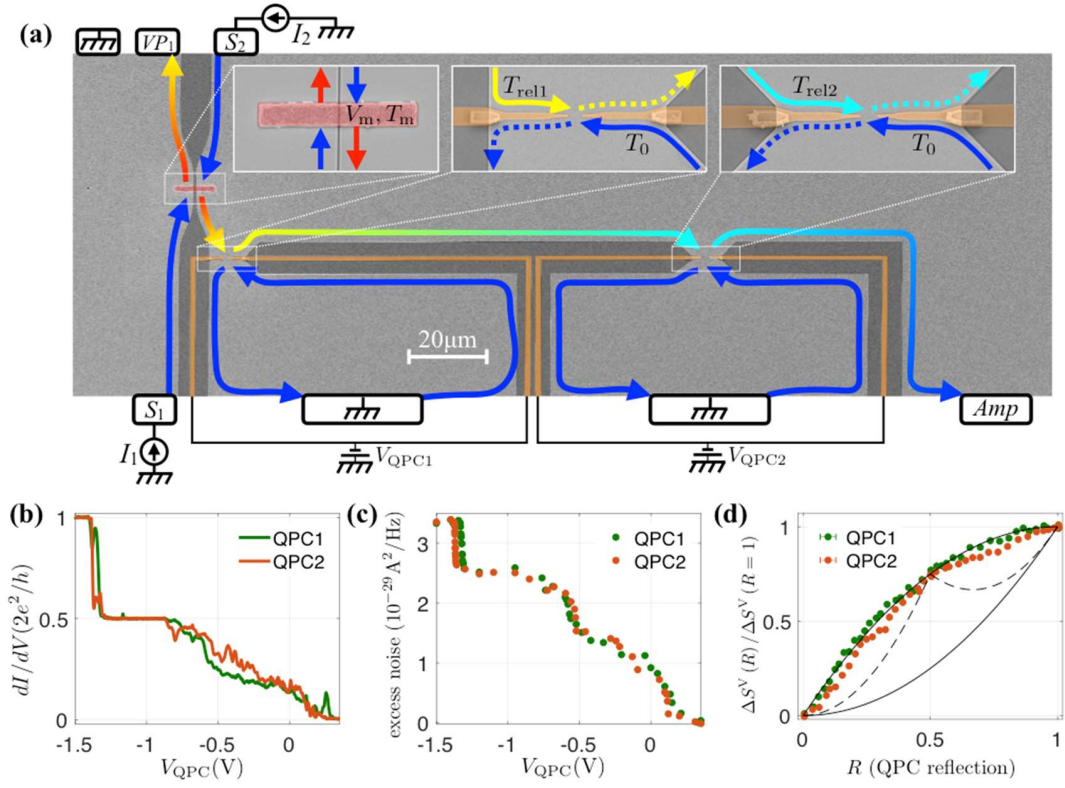

*Figure S3*

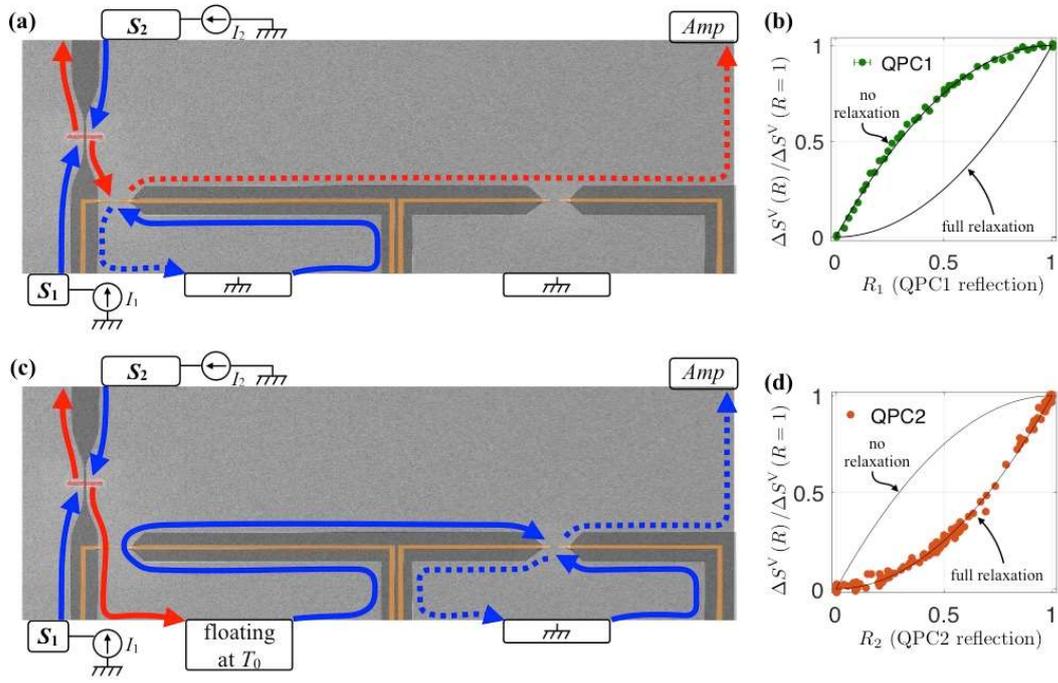

*Figure S4*

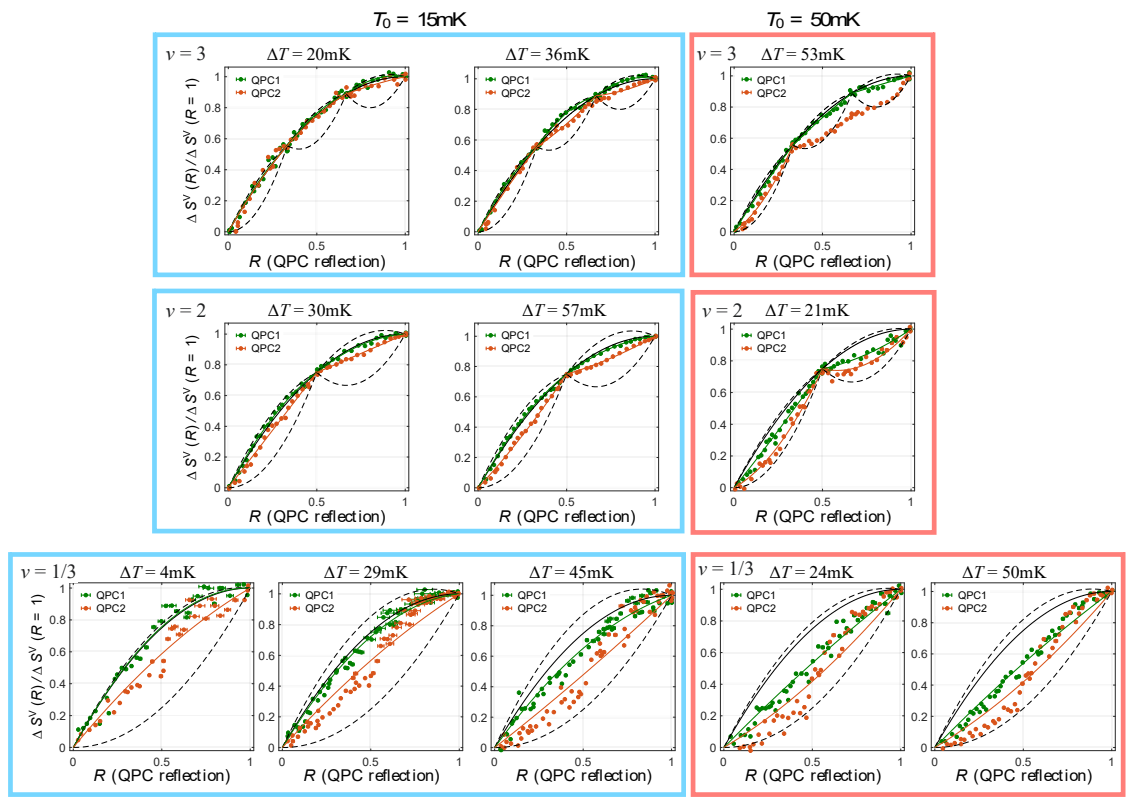

*Figure S5*

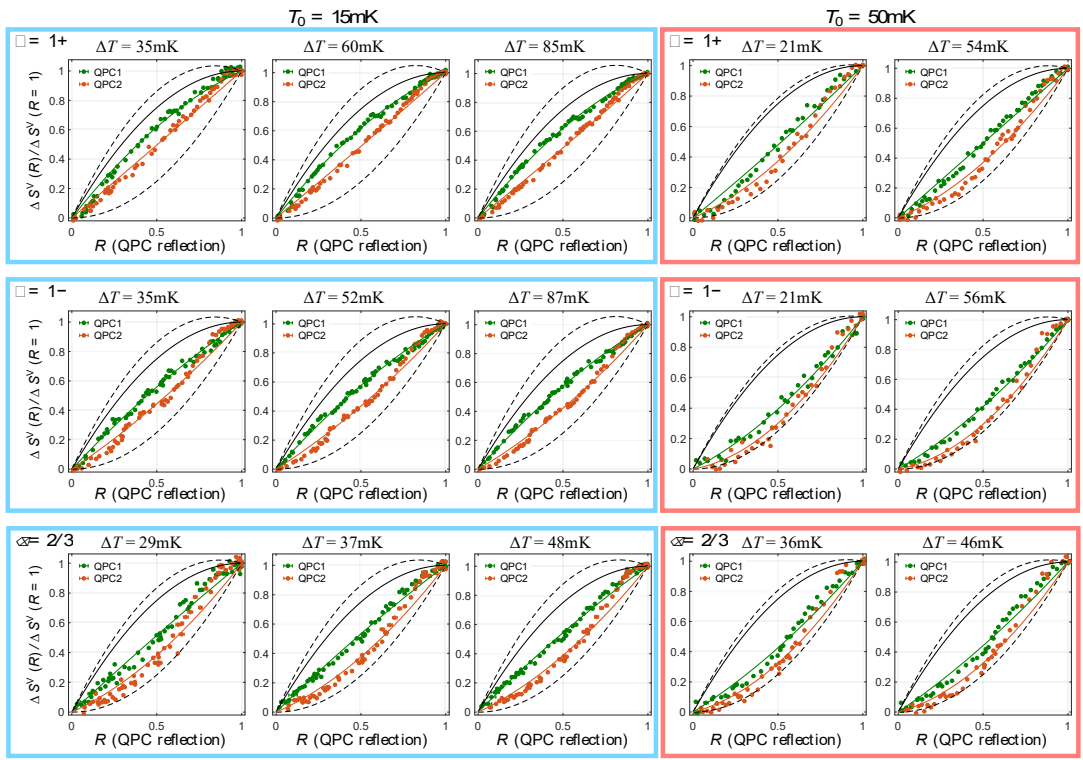

*Figure S6*

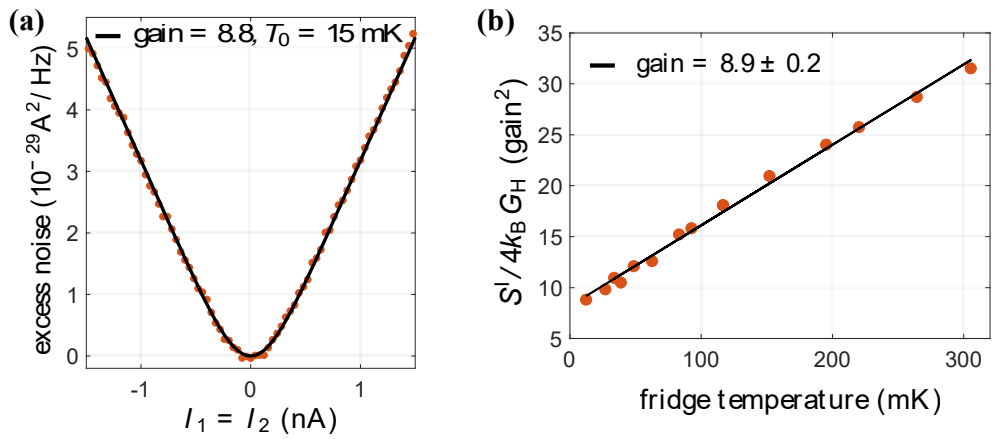

*Figure S7*

# Supplementary Figures Captions

Figure S1. **Temperature calibration of the small floating contact and the quantum limit of heat flow.** (a) Measurement of small floating contact's temperature $T_m$ as function of the heating current $I_1 = -I_2$, at base temperature $T_0 = 15$ mK. Measured when both QPCs are fully closed, at filling factors ν =3,2,1,2/3 and 1/3, where at ν =1 we measure on the low field side of the plateau (yellow) and on the high filed side (purple). (b) Temperature as function of the dissipated power $\Delta P$ for all measured filling factors, dashed black curves represents the theoretical quantum limit of heat flow for 2, 4 and 6 modes with electron-phonon term $\dot{Q}_{\text{el-p}} = \beta_{\text{el-p}} (T_m^5 - T_0^5)$ where $\beta_{\text{el-ph}} = 7\text{nW/K}^5$.

Figure S2. **Schematic description of the dissipation model.** Current flows from two sources at voltages $V_1, V_2$ and temperature $T_0$ to a floating contact. The floating contact is at voltage $V_m$ and temperature $T_m$, as described in the main text. Current leaves the floating contact via two arms, one of which is partitioned by a QPC en route to the Amp contact. A fictitious floating contact is placed right before the QPC. The fictitious floating contact has a temperature $T_{rel}$, which represents the effective local temperature of the edge at the point of partitioning, and a voltage $V_C$, which is determined self-consistently satisfying the demand that no charge accumulates within the contact.

Figure S3. **Partitioned thermal noise measurement methodology** (a) SEM image of the device, DC current with opposite polarity is sourced in $S_1$ and $S_2$ giving finite $\Delta P$ and $V_m = 0$. Edge modes flowing out from the right side of the small floating contact encounters two QPCs at 20μm and at 140μm away from the small floating contact. Conductance and noise are measured at the amplifier as function of the QPCs split-gate voltage. (b) Differential conductance (also reflection coefficient) of QPC1 and QPC2 as function of $V_{\text{QPC1}}$ and $V_{\text{QPC2}}$, measured at ν=2 (c) Excess noise as function of the split gate voltage of each QPC at ν=2. (d) Normalized noise as function of the reflection of each QPC, rendered from the two plots in (c) and (b).

Figure S4. **Forcing relaxation with cold floating contact.** (a) SEM image of the device, sourcing constant DC current from $S_1$ and $S_2$ with opposite polarity in order to heat the small floating contact and leave it with zero voltage. The hot channel is partitioned at QPC1, keeping QPC2 fully pinched, while noise is measured in the amplifier. (b) Normalized noise measured ν=2 as function of the reflection of the QPC1 as described in (a). The upper black curve is the non-relaxed limit and the lower black curves are the fully relaxed limit for a single channel. (c) Heating the small floating contact with constant DC current sourced from $S_1$ and $S_2$ with opposite polarity. The hot channel fully transmits through the first QPC and flows into a massive floating contact, cooling the edge down to base temperature $T_0$. The channel then flows to the second QPC and partitioned while measuring noise on the amplifier. (d) Normalized noise measured at ν =2 as function of the reflection of the second QPC as described in (c). The upper black curve is the non-relaxed limit and the lower black curve is the fully relaxed limit for a single channel.

Figure S5. **Thermal relaxation of edge modes in particle-like states.** Each plot shows the normalized noise as function of the reflection the QPC, while the other QPC is fully closed. On each plot we apply constant DC heating power ΔP and $V_m = 0$, raising the small floating contact's temperature to some $T_m$. At each filling factor we measure for few values of $T_m$ and for two base temperatures $T_0 = 15 mK$ and $T_0 = 50 mK$. Green data points are the normalized noise measured in QPC1, 20μm away from the small floating contact, and orange data points for QPC2, 140μm away from the small floating contact. The upper solid line is the no-relaxation limit for $h = 0$, the upper dashed line is the no-relaxation limit when taking into account $h(T_m/T_0)$, the lower dashed line is the full-relaxation limit (doesn't depend on $h$). We fit the noise with a relaxation parameter $\Theta$ plotted in green/orange solid lines. (a) $\Theta_{1\text{inner}} = 1.06$, $\Theta_{1\text{middle}} = 1.08$, $\Theta_{1\text{outer}} = 1.09$; $\Theta_{2\text{inner}} = 1.00$, $\Theta_{2\text{middle}} = 0.99$, $\Theta_{2\text{outer}} = 0.88$. (b) $\Theta_{1\text{inner}} = 1.11$, $\Theta_{1\text{middle}} = 1.16$, $\Theta_{1\text{outer}} = 1.12$; $\Theta_{2\text{inner}} = 0.91$, $\Theta_{2\text{middle}} = 0.73$, $\Theta_{2\text{outer}} = 0.74$. (c) $\Theta_{1\text{inner}} = 0.92$, $\Theta_{1\text{middle}} = 0.9$, $\Theta_{1\text{outer}} = 0.85$; $\Theta_{2\text{inner}} = 0.42$, middle and outer modes are mixed

at QPC2, no fitteing available. **(d)** $\Theta_{1\text{inner}} = 1.07, \Theta_{1\text{outer}} = 0.97; \Theta_{2\text{inner}} = 0.75, \Theta_{2\text{outer}} = 0.75$. **(e)** $\Theta_{1\text{inner}} = 1.05, \Theta_{1\text{outer}} = 0.94; \Theta_{2\text{inner}} = 0.63, \Theta_{2\text{outer}} = 0.67$. **(f)** $\Theta_{1\text{inner}} = 0.70, \Theta_{1\text{outer}} = 0.62; \Theta_{2\text{inner}} = 0.31, \Theta_{2\text{outer}} = 0.38$. **(g)** $\Theta_1 = 1.01, \Theta_2 = 0.68$. **(h)** $\Theta_1 = 0.94, \Theta_2 = 0.63$. **(i)** $\Theta_1 = 0.8, \Theta_2 = 0.46$. **(j)** $\Theta_1 = 0.57, \Theta_2 = 0.34$. **(k)** $\Theta_1 = 0.58, \Theta_2 = 0.34$.

Figure S6. **Thermal relaxation of edge modes in hole-conjugate states.** Each plot shows the normalized noise as function of the reflection the QPC, while the other QPC is fully closed. On each plot we apply constant DC heating power $\Delta P$ and $V_m = 0$, raising the small floating contact's temperature to some $T_m$. At each filling factor we measure for few values of $T_m$ and for two base temperatures $T_0 = 15\text{mK}$ and $T_0 = 50\text{mK}$. Green data points are the normalized noise measured in QPC1, 20µm away from the small floating contact, and orange data points for QPC2, 140µm away from the small floating contact. The upper solid line is the no-relaxation limit for $h = 0$, the upper dashed line is the no-relaxation limit when taking into account $h(T_m/T_0)$, the lower dashed line is the full-relaxation limit (doesn't depend on $h$). We fit the noise with a relaxation parameter $\Theta$ plotted in green/orange solid lines. **(a)** $\Theta_1 = 0.74, \Theta_2 = 0.49$. **(b)** $\Theta_1 = 0.74, \Theta_2 = 0.49$. **(c)** $\Theta_1 = 0.74, \Theta_2 = 0.47$. **(d)** $\Theta_1 = 0.46, \Theta_2 = 0.24$. **(e)** $\Theta_1 = 0.50, \Theta_2 = 0.24$. **(f)** $\Theta_1 = 0.62, \Theta_2 = 0.39$. **(g)** $\Theta_1 = 0.65, \Theta_2 = 0.37$. **(h)** $\Theta_1 = 0.65, \Theta_2 = 0.35$. **(i)** $\Theta_1 = 0.29, \Theta_2 = 0.10$. **(j)** $\Theta_1 = 0.34, \Theta_2 = 0.10$. **(k)** $\Theta_1 = 0.53, \Theta_2 = 0.26$. **(l)** $\Theta_1 = 0.57, \Theta_2 = 0.25$. **(m)** $\Theta_1 = 0.59, \Theta_2 = 0.28$. **(n)** $\Theta_1 = 0.35, \Theta_2 = 0.15$. **(o)** $\Theta_1 = 0.39, \Theta_2 = 0.15$.

Figure S7. **Calibration of the amplifier. (a)** Shot-noise measurement at $\nu = 2$ for calibration of gain and temperature. **(b)** Thermal noise as function of base temperature of the fridge to verify calibration of gain at $\nu = 2$.